# A perspective on Microscopy Metadata: data provenance and quality control

Maximiliaan Huisman (https://orcid.org/0000-0003-3363-8791)[1], Mathias Hammer[1], Alex Rigano[2], (https://orcid.org/0000-0001-7471-2244) Ulrike Boehm[#3], (https://orcid.org/0000-0003-3883-8215) James J. Chambers[#4], (https://orcid.org/0000-0002-9220-5366) Nathalie Gaudreault[#5], (https://orcid.org/0000-0002-0577-7949) Alison J. North[#6], (https://orcid.org/0000-0001-8569-0466) Jaime A. Pimentel[#7], (https://orcid.org/0000-0002-2510-7272) Damir Sudar[#8], (https://orcid.org/0000-0002-6968-2615) Peter Bajcsy[#9], (https://orcid.org/0000-0003-1622-663X) Claire M. Brown[#10], (https://orcid.org/0000-0003-1645-5475) Alexander D. Corbett[#11], (https://orcid.org/0000-0001-5965-5405) Orestis Faklaris[#12], (https://orcid.org/0000-0002-8783-8599) Judith Lacoste[#13], (https://orcid.org/0000-0002-3853-1187) Alex Laude[#13], (http://orcid.org/0000-0002-1895-4772) Glyn Nelson[#14], (https://orcid.org/0000-0002-9397-8475) Roland Nitschke[#15], (https://orcid.org/0000-0001-9067-804X) David Grunwald[1], and (https://orcid.org/0000-0002-1069-1816) Caterina Strambio-De-Castillia[#2]

# Members of the Bioimaging North America Quality Control and Data Management Working Group

[1] RNA Therapeutics Institute, UMass Medical School, Worcester MA 01605, USA
[2] Program in Molecular Medicine, UMass Medical School, Worcester MA 01605, USA
[3] Janelia Research Campus, Howard Hughes Medical Institute, Ashburn, VA 20147, USA
[4] Institute for Applied Life Sciences, University of Massachusetts, Amherst, MA 01003, USA
[5] Allen Institute for Cell Science, Seattle, WA 98109, USA
[6] The Rockefeller University, New York, NY 10065, USA
[7] Instituto de Biotecnología, Universidad Nacional Autonoma de Mexico, Cuernavaca, Morelos, 62210, México
[8] Quantitative Imaging Systems LLC, Portland, OR 97209, USA
[9] National Institute of Standards and Technology, Gaithersburg, MD 20899, USA
[10] Advanced BioImaging Facility (ABIF), McGill University, Montreal, Quebec, H3G 0B1, Canada
[11] Department of Physics and Astronomy, University of Exeter, Exeter, EX4 4QL, UK
[12] Biocampus, CNRS UMS 3426, Montpellier 34293, France
[13] MIA Cellavie Inc., Montreal, Quebec, H1K 4G6, Canada
[14] Bioimaging Unit, Newcastle University, Newcastle upon Tyne, NE2 4HH, UK
[15] Life Imaging Center and BIOSS Centre for Biological Signaling Studies, Albert-Ludwigs-University Freiburg, Freiburg, 79104, Germany

# 1 - SUMMARY

The application of microscopy in biomedical research has come a long way since Antonie van Leeuwenhoek discovered unicellular organisms (or *dierken*s in his original letters and subsequently translated in English as *animalcules*) (*7*, *8*) through his hand-crafted microscope. Countless innovations have propelled imaging techniques and have positioned light microscopy as a cornerstone of modern biology and a method of choice for connecting 'omics' datasets to their biological and clinical correlates. Still, regardless of how convincing published imaging data *looks*, it does not always convey meaningful information about the conditions in which it was acquired, processed, and analyzed (*9–16*). Adequate record-keeping, reporting, and quality control are therefore essential to ensure experimental rigor and data fidelity, allow experiments to be reproducibly repeated, and promote the proper evaluation, interpretation, comparison, and re-use of the results. Microscopy images should be accompanied by complete descriptions detailing experimental procedures, biological samples, microscope hardware specifications, image acquisition parameters and image analysis procedures, as well as metrics accounting for instrument performance and calibration. However, despite considerable efforts (*10*, *17*, *18*), universal, community-accepted Microscopy Metadata standards and reporting specifications for the Findable Accessible Interoperable and Reproducible (FAIR; Text Box I) (*19*), assessment, reproducibility, sharing and comparison of microscopy data have not yet been established. To understand this failure and to propose a way forward, here we provide an overview of the nature of microscopy metadata and its importance for fostering data quality, reproducibility and scientific and sharing value (i.e., the extent to which the data serves its intended scientific purpose and can be shared with other scientists to extract further insight) in light microscopy. In order to center the discussion, in this paper we consider a light microscopy image as "a digital representation of an optical image of the distribution of endogenous or introduced contrast in the specimen" (modified from: (*20*)).

# 2 - METADATA

Metadata is data that provides information about other data. In addition to basic metadata, such as the name and date of a file, rich metadata consists of any and all additional information that allows data to be found, evaluated, cited, reproduced, and re-used by third-party users and (ideally) by automatic means. In general, rich metadata adds value to the measured data by improving the degree to which its fidelity can be ascertained and its ability to meet quality criteria (i.e., FAIR principles) (*19*) can be assessed. In general, many distinct categories of metadata exist. Basic metadata categories include descriptive metadata (e.g., file name, description, and author), and administrative metadata (e.g., data owner, time of creation). Rich metadata categories are more variable and depend strongly on the specific requirements of individual disciplines. In the case of imaging experiments, it is possible to distinguish between Image Metadata (Table I) and Microscopy Metadata (Tables II and III) (*2*, *3*, *10*). In particular, Image Metadata consists of any and all information that allows imaging results to be evaluated, interpreted, reproduced, found, cited, and re-used as established by measurable data quality criteria (i.e., FAIR principles) (*19*). As such, Image Metadata can be defined as metadata that documents all phases of a typical bioimaging experiment including sample preparation and labeling, image acquisition, and image processing and analysis. On the other hand, we define Microscopy Metadata as the subset of Image Metadata that specifically concentrates on the process of Image Acquisition using a Microscope and the Quality of the resulting Image Data. As such, Microscope Metadata can be subdivided into two sub-categories: 1) Microscopy data "Provenance" Metadata (MPM) includes microscope hardware specifications, image acquisition settings and image structure; and 2) Microscopy Quality-control Metadata (MQM) includes calibration metrics that quantitatively assess the performance of the microscope. Together, MPM and MQM maximize the repeatability, reproducibility, accuracy, and precision of results, establish whether the conclusions are warranted, and enable comparisons both within and between datasets. In general, the more Image and Microscopy metadata is provided, the higher the image data quality, reproducibility, and scientific and sharing value, and the easier it is to develop appropriate image analysis strategies. Despite the importance of both Image and Microscopy Metadata, for the purpose of this discussion we concentrate primarily on Microscopy Metadata (Tables II and III).

## 2.1 - Microscopy Data Provenance Metadata: Describing the origin of the data and how it was produced

'Data provenance,' also referred to as data "lineage" or "pedigree," is an overloaded phrase with multiple definitions (*21*). In essence, 'data provenance' consists of an electronic version of the scientific record (lab notebook) that is routinely maintained alongside experiments to document all steps of the procedure (experimental setup, sample preparation, instrument characteristics and settings, quality control procedures, analysis algorithms, and parameters). Hence, data provenance extends the basic scientific requirement of thorough experimental documentation to the domain of "big-data." Because the sheer amount of data makes manual annotation impractical, data provenance increasingly relies on computational processes that are obscure to the experimenter, such as data acquisition steps, data formatting, data processing, as well as algorithm input parameters and source code. Thus, all information pertaining to the origin of the data needs to be electronically logged, recorded, and maintained in both a human- and a machine-readable manner to ensure data quality, foster its reproducibility and evaluation and maximize its scientific value.

To formally define data provenance, Ram and Liu proposed an ontological model called W7 (Figure 1), which might be used to track events that affect data during its lifetime (*22*). The W7 model defines provenance as the ensemble of metadata that facilitate scientific reproducibility, repeatability, validation, and interpretation by: (1) describing "what" happened to each data element (i.e., data manipulation events such as acquisition, processing, and analysis); and (2) providing answers to six interconnected "who," "how," "when," "where," "which," and "why" questions describing each data manipulation event. Following the W7 model, there are four aspects of data provenance that are generally considered:
1. Tracing the origin of each piece of raw, intermediate, and final data.
2. Recording information about scientific experiments and sample preparation.
3. Keeping track of data processing, analysis, and visualization.
4. Annotating data with 1, 2 & 3.

The types of MPM that need to be recorded in a typical light microscopy experiment are listed in Table II.

### 2.1.1 - Describing the instrument

An intimate knowledge and detailed record of the specific microscopy hardware used for the experiment is just as important as detailed documentation of the sample and experimental design. Indeed, the availability and compatibility of specific hardware components, as well as the overall state of the equipment, play an essential role in determining the experimental outcome and in defining to what extent the resulting images can be utilized to answer a given scientific question. The following important categories of microscope hardware information will impact the results:

- **Microscope hardware specifications:** The effect of optical, electronic, and mechanical design upon image formation and interpretation increases with instrument complexity. Knowledge about the light source type, wavelength, and intensity, and about the optical path taken by light on its way to the sample and then to the detector, can help identify effects like crosstalk between color channels, high background fluorescence, and non-linear photo-physical effects. For example, information about the detector aids in understanding both image quality and the choice of quantitative analysis strategies to be used. It is therefore imperative to detail the physical characteristics of the microscope and of all other components that were utilized during the image formation and acquisition process (e.g., microscope manufacturer and model number, nominal objective magnification, numerical aperture and performance class - according to ISO, filter excitation and emission wavelength, etc.), as well as the software and version used to drive image acquisition.

### 2.1.2 - Describing image acquisition

The design and execution of biomedical imaging experiments involve many decisions, from the choice of imaging technique (e.g., fluorescence vs. transmitted light), sample labeling (e.g., fluorophore and molecular target), to the exposure time of a camera. In turn, these choices affect the interpretation of results and the type of measurements that can be performed on an image. Categories of metadata related to these choices include:

- **Administrative information:** In order to track potential confounder and in particular batch effects that might affect the interpretation of results (*15*), it is essential to keep a permanent record of when, where, and by whom the images were acquired.

- **Minimal sample information:** Not all details about sample preparation need to be included as Microscopy Metadata. However, key details about sample preparation that are required for reproducibility, meaningful data interpretation, as well as for selecting the most appropriate imaging technique, include the labeling technique, the fluorophores used for visualization, and any special sample preparation, fixation, or mounting methods.

- **Sample labeling:** While an intimate knowledge of the technique used to create enhanced sample contrast is important for all modalities, this information is particularly relevant in the case of fluorescence microscopy because fluorescent molecules can only be efficiently excited within a specific wavelength range. Thus an intimate knowledge of the physical properties (e.g., stability, absorption, quantum yield) of the specific fluorescent dye used in the experiment is an essential starting point for ensuring the reproducibility of the experiment. More specifically, the brightness of the fluorescence signal is proportional to the excitation power, but so are the photobleaching rate and the potential of light-induced damage to the cell. Hence, using appropriate fluorescent dyes, excitation wavelengths and intensity is crucial both for illuminating the fluorescent molecules effectively and for avoiding light-induced artifacts.

- **Imaging technique:** Individual imaging modalities are commonly associated with specific types of data artifacts. More specifically, different imaging techniques might differ in terms of excitation method (e.g., wide-field, confocal, TIRF, light-sheet, etc.), signal detection (e.g., cameras vs. point-detectors), and acquisition time ranging from picoseconds to minutes. In turn, these differences might imply variations in sample preparation, light intensity or noise characteristics, which in turn result in artifacts that depend on individual techniques. For example, structured illumination or single-molecule super-resolution techniques often lead to artifacts in reconstructed or processed data, while light-intensive microscope modalities pose a high risk of biological impact on the sample through phototoxicity. In any case, the confident interpretation of the results requires a thorough understanding of the pros and cons of the chosen imaging method.
- **Microscope configuration:** Microscopes are often customized and upgraded over time and can be configured in various ways. The choice of optical components (e.g., objective lenses, emission filters, and dichroic mirrors) can dramatically affect the acquired images, especially if they are not ideally suited for the experiment. Similarly, the invocation of options like scaling adapters, optical relays, or mechanical stabilizers (e.g., drift-correction and autofocus) can affect data interpretation by altering microscope performance.
- **Image acquisition settings:** Some information (e.g., the need to match the refractive index of the mounting medium with the lens immersion liquid) is essential for even the most basic microscopy data because it influences resolving power and scale. For more complex experiments (e.g., three-dimensional, time-lapse and multi-point acquisition, or low-signal conditions), additional parameters (e.g., optical sectioning distance, frame rate, and exposure time) must also be reported. In general, more sophisticated hardware is associated with more parameters to be established and reported. For example, when using a highly sensitive Electron-Multiplying Charge-Coupled Device (EMCCD) camera, the EM-gain and sensor temperature must be recorded in addition to the image exposure time and pixel size.

### 2.1.3 - Describing the structure of the Image

The data structure of a digital image is often multidimensional and quite complex. Thus, clear and well-ordered information indicating how the digital structure of the images should be interpreted (pixel size, number of dimensions utilized, which dimensions constitute different focal planes, color-channels, etc.) is crucial:

- **Image-structure information:** The basic digital image structure consists of planes each composed of a two-dimensional matrix of pixels (X and Y dimensions). In addition, modern microscopy images commonly consist of many planes obtained under different illumination conditions (color-channels, C dimension), over multiple focal planes (Z dimension), timeframes (T dimension), and more advanced dimensions such as/or spectral conditions ($\lambda$ dimension), polarization angles, etc.

## 2.2 - Microscopy Quality Control Metadata: Measuring instrument performance, estimating error, and reporting quality-control metrics

The process of image formation in optical microscopy can be subdivided into multiple steps, each of which is a potential source of uncertainty. Thus, each image formation step must be documented and its impact on error propagation quantified to determine how accurately the resulting image represents the biological phenomenon being studied. In this context MQM represents all metadata that adds rigor and fidelity to the data. MQM typically comprises metrics related to the quantification and calibration of instrument performance and estimation of error propagation (see Table II).

Ideally, every measurement obtained using the light detector associated with a given microscope would be consistent with its stated performance specifications, minimize the amount of bias and imprecision attributable to the instrument, and be readily interpretable. In reality, microscopes are complex assemblies that require frequent tuning and calibration, careful handling, and well-trained users who understand the limits to which they should interpret the data obtained. Even well-maintained microscopes may have multiple users performing different kinds of experiments that require settings to be altered or filters and optics to be exchanged, which can cause accidental misalignment or soiling. To make matters worse, subtle - even imperceptible - performance defects can dramatically affect imaging results and, therefore, the interpretation of an experiment. Whereas obvious artifacts may readily raise the alarm, subtle defects will only be detected by careful measurement. This is especially true for data that requires quantitative analysis and/or pushes the technical limits of the microscope since this is where the more subtle effects of sub-optimal calibration may be revealed. To facilitate the repeatability, reproducibility, and comparison of experimental results arising from imaging experiments - even long after images were acquired - image datasets must be annotated with detailed, quantitative measurements of the microscope performance obtained at the time of acquisition.

*2.2.1 - Measuring instrument performance and calibration*

Since microscopes with identical components (e.g., light source, optics, camera) can show significant performance variations, standardized performance evaluations undertaken in advance of an experiment would reduce the need for expensive and time-consuming experimental repetitions. They would also allow for problems to be identified and resolved before they become significant issues. Even if performed later, such measurements would help the researcher determine how the data can be interpreted, and the extent to which it can be used to answer the scientific question of interest, as well as help identify potential batch effects. The use of various metrics has been proposed to measure microscope performance depending on the type and intent of the experiment (*20, 23–45*). However, there is a general lack of consensus on what quality control metrics, procedures, and tools should be used in each case (*46*). Either way, the categories of metrics about the microscope performance listed in Table II and discussed below are important to consider as they impact the results:

- **Quality control metrics:** Calibration metrics can be generally subdivided into four categories: (1) optical; (2) intensity_excitation; (3) intensity_detector; (4) mechanical (Table II). Metrics in categories 1-3 evaluate the great majority of image data. Mechanical calibration metrics become most useful in experiments that involve single or multi-position time-lapse imaging or the tiling of multiple Fields-Of-View (FOV) when imaging large samples. Together, these measurements increase the depth and reliability of various assessments, analyses, and comparisons performed on light microscopy images.

## 2.3 - Ideal properties of microscopy metadata standard guidelines

The absence of universally accepted metadata standards precludes the full realization of quantitative microscopy's potential to become a reliable source of unbiased and reproducible quantitative data. We propose that a realistic and attainable metadata standard should strive towards ideal criteria (listed below), which comply with the emerging Minimum Information (*47*) and FAIR (*19*) principles for data stewardship:

1. **Metadata and data should be easy to comprehend and find by experimental scientists.** The terms utilized by the standard need to be self-explanatory, domain-relevant, broadly applicable, generally understood by the community, and easy for microscopy users to interpret.

2. **Microscopy data and metadata should be easy for machines to access and interpret.** Machine-readable metadata is essential to maximize automation, thereby minimizing the dataset annotation overhead and redundancy for experimental scientists, as well as facilitating searchable resources. For the latter, datasets should be unequivocally identifiable, and metadata should contain explicit links to the data they describe.

3. **Formal data model.** Metadata should be structured in a formal data model describing the real-life components that need to be documented, the relationship between components, and the attributes that need to be recorded to sufficiently describe each component. Furthermore, the model must support extensions to accommodate the addition of new components, relationships, and attributes.

4. **Reusability.** To optimize the reuse of data, metadata and data should be well-described so that they can be replicated, compared, and combined with data obtained by other scientists under comparable conditions. The standard should clearly stipulate a <u>rich plurality of relevant metadata parameters</u> to describe the data, specify acceptable units of measure, and guarantee easy access and navigation.

5. **Interoperability.** Data and metadata should be easy to integrate with other data and readily accessible with available applications or workflows for analysis, storage, and processing. For this purpose, the standard should include strong links that connect data and metadata to other relevant data and metadata.

6. **Tracing the origin of data and all intermediate steps.** The standard should stipulate that all necessary information to trace the 'provenance' (origin, source, interlinking) of data (e.g., microscope set-up, image acquisition settings, and image analysis software used) should be documented in detail.

7. **Maximal usability and minimal annotation burden**

    a. **Minimal information:** The standard should require <u>only</u> those parameters that can be reasonably expected to influence the outcome of the microscopy experiment.

    b. **Adaptable system of guidelines:** Recognizing that different experiments often have different annotation requirements, the standard should follow an adaptable system of norms that scales with experimental and technological complexity and that specifies which data provenance and quality control metadata should be provided for a given case, thus minimizing the documentation burden on the experimenter.

8. **Standardized instrument calibration and error estimation.** The standard should provide explicit norms about recommended quality control procedures and about the frequency with which each should be performed to ensure experimental rigor and data fidelity.

## 2.4 - How to collect comprehensive metadata annotation and obtain standardized metrics for instrument calibration

The importance of rich metadata to ensure the quality, reproducibility, as well as scientific and sharing value of image data cannot be overstated (*9*, *46*, *48–51*). As a first step to address this need, the 4D Nucleome initiative (*52*, *53*) and Bioimaging North America (BINA) (*54*, *55*) have joined forces to develop tiered Microscopy Metadata Specifications that extend the OME Data Model (*1–3*) and provide standardized specifications for both MPM and MQM entries. However, the collection of rich sets of provenance and quality control metadata is time-consuming, complex and costly. In addition, in the absence of active participation from hardware manufacturers, standards impose an unfair burden on experimental scientists and are therefore difficult to enforce.

In this context, appropriate community-validated software tools and data management practices (*56*, *57*) are essential to streamline and automate the collection, storage, and management of MPM (Table II) information needed to appropriately document microscopy experiments. Within this framework, a suite of three complementary and interoperable tools are being developed to facilitate the process of image data documentation and are presented in related manuscripts.

(1) OMERO.mde (*4*) focuses on two aspects: 1) facilitating the consistent handling of image metadata ahead of data publication and deposition based on shared community Microscopy Metadata specifications and according to the FAIR principles; 2) the early development of Image Metadata extension specifications to maximize flexibility while at the same time allowing for testing and validation before incorporation in community-accepted standards.

(2) Micro-Meta App (*6*, *58*) focuses on an easy-to-use, graphical user interface (GUI)-based platform that allows users to interactively build models of microscope hardware, accessories and image acquisition settings containing all relevant Microscopy Metadata as sanctioned by the community (*2*, *3*). Because of its graphical nature, Micro-Meta App is particularly suited for teaching trainees about Microscopy, and training microscope-users in facilities.

(3) Finally, MethodsJ2 (*5*) focuses on capturing Microscopy Metadata that should be reported in publications and automates the process of writing appropriate Methods and Acknowledgment sections to maximize efficiency, quality, reproducibility, and scientific and sharing value. MethodsJ2, by design, operates in concert with the Micro-Meta App to automatically import Microscopy Metadata from the JSON files produced using the App.

Different calibration metrics (Table III) require different measurement procedures, each of which has unique advantages and disadvantages (*20*, *23–36*). To make things even more complex, calibration methods typically require often costly equipment (e.g., reference slides, power meters, calibrated light-sources) and entail levels of skill and time commitment that cannot be reasonably expected from most microscope users. The absence of clear and shared guidelines to define intuitive, repeatable, and inexpensive quality control routines to standardize the calibration of fluorescence microscopes has recently prompted the launch of the **QU**ality Assessment and **REP**roducibility for Instruments and Images in Light Microscopy (QUAREP-LiMi) initiative with the explicit aim of reaching a broad consensus to be adopted worldwide (*46*, *51*). In parallel, the Meta-Max hardware tool was developed to standardize detector and excitation calibration, of particular use for image data with a low Signal-to-Noise Ratio (SNR) and is subject to statistics-based image analysis (*59*, *60*).

Finally, all the metadata needs to be stored and closely tied to the image pixel values in easy-to-use containers (e.g., files or other organized storage structures) that are highly standardized, provide high-performance access methods, can be shared easily, and whose design ensures their longevity. The OME-led Next Generation File Format (NGFF) (*61*, *62*) initiative is developing such container(s) and is working closely with the above-mentioned community groups.

## 3 - CONCLUSIONS

We acknowledge that the achievement of perfect standards in light microscopy may be a considerable challenge. Nevertheless, a system of clear guidelines based around minimal requirements and sensible compromises would greatly enhance the currently poor status of rigor and reproducibility for light microscopy data, and would help the field mature in ways that benefit the entire life sciences community. In order to stimulate such progress, in our accompanying manuscript, we delineate a set of scalable light microscopy metadata specifications that extend the OME Data Model (*17*, *63*), are compatible with the recently proposed minimum recommended metadata for reuse of imaging data (*56*) and have emerged from work conducted by the Imaging Standard Working Group (IWG) of the National Institutes of Health (NIH) Common Fund's 4D Nucleome (4DN) Consortium (*52*, *53*), in collaboration with the Quality Control and Data Management Working group of BioImaging North America (BINA) (*55*) and with QUAREP-LiMi (*46*, *51*).

## 4 – AUTHORS CONTRIBUTIONS

Author contributions categories utilized here were devised by the CRediT initiative (*64*, *65*).
**Mathias Hammer**: Conceptualization, Methodology, Investigation, Data Curation; **Maximiliaan Huisman**: Conceptualization, Methodology, Investigation, Writing - Original Draft, Writing - Review & Editing, Visualization; **Alessandro Rigano**: Conceptualization, Methodology, Software; **Ulrike Boehm**, **James J. Chambers**, **Nathalie Gaudreault**, **Alison J. North**, **Jaime A. Pimentel**, and **Damir Sudar**: Validation, Data Curation, Writing - Review & Editing; **Peter Bajcsy**, **Claire M. Brown**, **Alexander D. Corbett, Orestis Faklaris, Judith Lacoste**, **Alex Laude**, **Glyn Nelson**, and **Roland Nitschke:** Validation, Writing - Review & Editing; **David Grunwald**: Conceptualization, Methodology, Investigation, Resources, Writing - Original Draft, Supervision, Project administration, Funding acquisition; **Caterina Strambio-De-Castillia**: Conceptualization, Methodology, Software, Validation, Investigation, Resources, Data Curation, Writing - Original Draft, Writing - Review & Editing, Visualization, Supervision, Project administration, Funding acquisition.

## 5 - ACKNOWLEDGEMENTS

We would like to thank Kevin Fogarty, Lawrence Lifshitz and Karl Bellve at the Biomedical Imaging Group of the Program in Molecular Medicine at the University of Massachusetts Medical School for invaluable intellectual input and countless fruitful discussions and for their friendship, advice, and steadfast support throughout the development of this project.

This project could never have been carried out without the leadership, insightful discussions, support and friendship of all OME consortium members, with particular reference to Jason Swedlow, Josh Moore, Chris Allan, Jean Marie Burel, and Will Moore. We are massively indebted to the RIKEN community for their fantastic work to bring open science into biology. We would like to particularly acknowledge Norio Kobayashi and Shuichi Onami for their friendship and support.

We thank all members of Bioimaging North America (in particular Lisa Cameron, Michelle Itano, and Paula Montero-Llopis), German Bioimaging (in particular Susanne Kunis and Stephanie Wiedkamp Peters), Euro-Bioimaging (in particular Antje Keppler and Federica Paina) and QUAREP-LiMi (in particular all members of the Working Group 7 - Metadata; quarep.edu) for invaluable intellectual input, fruitful discussions and advice. We are also indebted to the following individuals for their continued and steadfast support: Jeremy Luban, Roger Davis, and Thoru Pederson at the University of Massachusetts Medical School; Burak Alver, Joan Ritland, Rob Singer, and Warren Zipfel at the 4D Nucleome Project; Ian Fingerman, John Satterlee, Judy Mietz, Richard Conroy, and Olivier Blondel at the NIH. We would like to thank Dr. Darryl Conte for the critical reading of the manuscript.

This work was supported by NIH grant #1U01EB021238 and NSF grant 1917206 to D.G., NIH grant # 2U01CA200059-06 to C.S.D.C and D.G., and by grant #2019-198155 (5022) awarded to C.S.D.C. by the Chan Zuckerberg Initiative as part of their Imaging Scientist Program. C.M.B. was funded in part by grant #2020-225398 from the Chan Zuckerberg Initiative DAF, an advised fund of Silicon Valley Community Foundation. D.S. was funded in part by NIH/NCI grants U54CA209988 and U2CCA23380. R.N. was funded by the Deutsche Forschungsgemeinschaft (DFG, German Research Foundation) grant number Ni 451/9-1 MIAP-Freiburg.

# TEXT BOXES

## Text Box 1

### FAIR data principles for scientific data management and stewardship

The **FAIR data principles for scientific data management and stewardship** were developed by a diverse set of stakeholders — representing academia, industry, funding agencies, and scholarly publishers — and published on Scientific Data in 2016 (*19*). The purpose of these principles is to enhance the reproducibility, reusability and value of scientific output. As a result of the increase in volume, complexity, and creation speed, scholars increasingly rely on computational support to interact with digital data assets. Thus, the FAIR Principles put specific emphasis on enhancing the ability of machines to automatically find and use the data, in addition to supporting its reuse by individuals. The FAIR guidelines state that in order to maximize reusability value, digital data should be: Findable, Accessible, Interoperable. A detailed description of the significance of each of these principles can be found online (*66*). A brief summary is listed below:

### **F**indable
The first step in (re)using data is to find them. For this purpose, metadata and data should be easy to find for both humans and computers and should be machine-readable.

### **A**ccessible
Once the required data has been located, it has to be easy to access via clear authentication and authorization procedures.

### **I**nteroperable
Easy to use applications and workflows have been commonly available to be able to store, process and analyze the data. In addition, data usually need to be integrated with other data.

### **R**eusable
To maximize data reusability metadata and data should be well-described and documented so that they can be replicated and/or combined with comparable data produced in different settings.

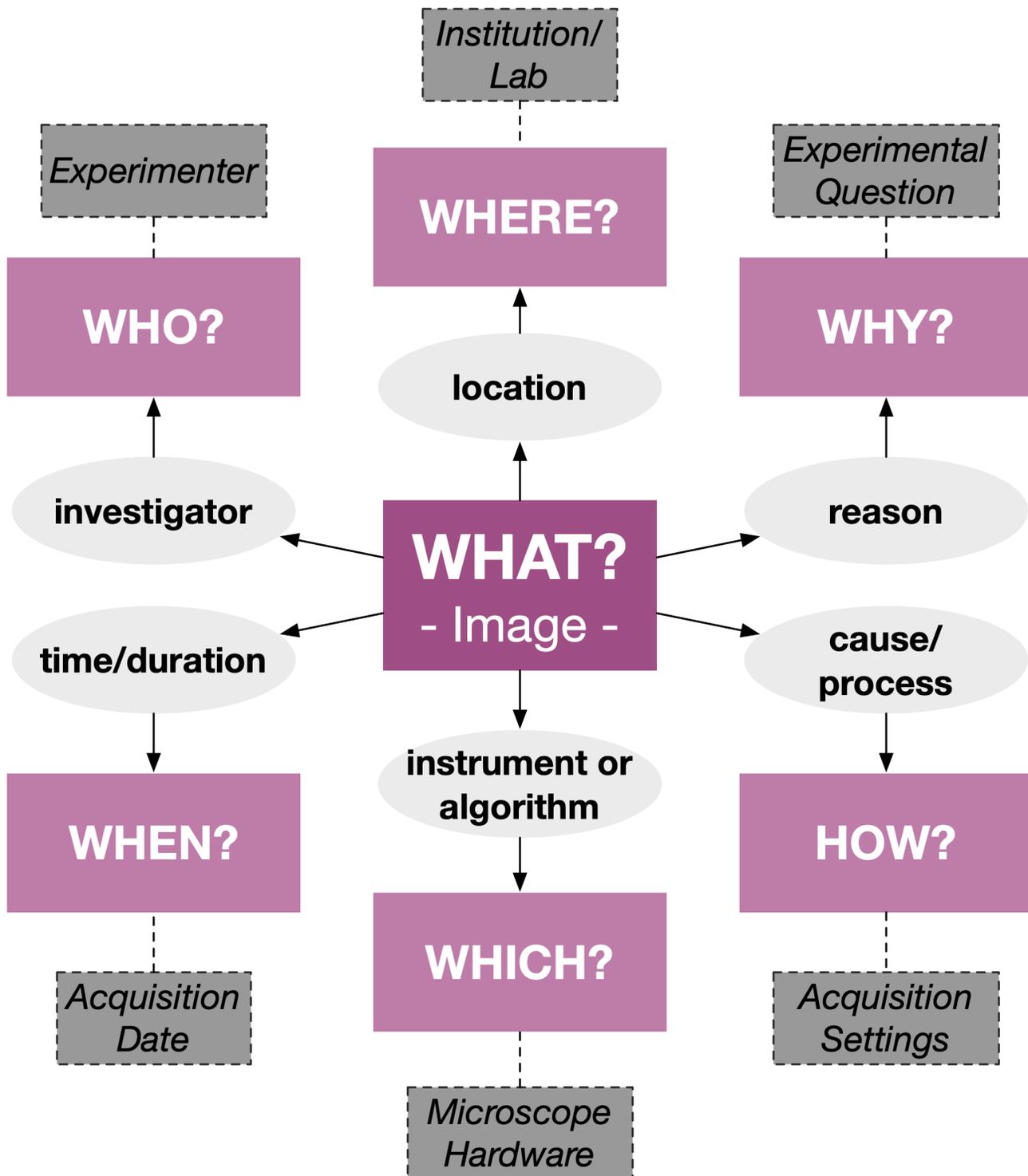

**Figure 1: W7 model of data provenance as applied for Microscopy Metadata.** Depicted is a conceptual graph representing the W7 model of data provenance devised by Ram and Liu, 2009. In this graph, the WHAT? in the central dark pink box represents an event/object/datum whose provenance needs to be documented, which in this case corresponds to an individual Image. Lighter pink boxes surrounding the central box represent questions about an individual Image, which are answered by individual pieces of metadata. Grey circles represent the kind of information that might answer individual provenance questions. Pink dashed boxes represent example metadata keys for documenting data provenance in the case of Microscopy Metadata. Adapted from Ram and Liu, 2009.

# Table I

**Table I: Image Metadata describes all aspects of an imaging experiment including the process of Image Acquisition**

| Category | Explanation of metadata | Examples | Collection method | Standards |
|---|---|---|---|---|
| Experimental metadata | Captures all experimental setup information that would be routinely recorded in a lab notebook | Tissue culture conditions, number of conditions, number of biological replicates | Mostly manual data entry | Domain-specific. Example: MIBBI, MISFISHIE; MIAME, CSMD; Fbbi [1] |
| Sample preparation | Documents all sample preparation information that would be routinely recorded in a lab notebook | Fixation and permeabilization conditions, staining conditions, mounting conditions | Mostly manual data entry | Domain-specific. Example: BRISQ; Fbbi [2] |
| Image Acquisition | See Tables II and III | | | |
| Data analysis | Describes all steps of the image processing, analysis, and visualization workflow | Algorithm name, algorithm version, algorithm parameters | In principle, easy to automatically capture, log and maintain during analysis; in practice, rarely achieved | Example: MIAPTE; MIACME [3] |
| Data dissemination | The information that is required for publication and data dissemination | Submission title, submission identification, submission description | Easy to automatically capture | Example: ISAtab; SampleTab [4] |

**Legend:** 1 - (48,67–73). 2 - (69,74). 3 - (75–81). 4 - (82–84).

# Table II

| Table II: Microscopy Metadata - Data Provenance Metadata | | | | |
|---|---|---|---|---|
| *Category* | *Explanation of metadata* | *Examples* | *Collection method* | *Standards* |
| **Microscope Hardware Specifications** | Describes the microscope used for image acquisition and lists its hardware components | Objective manufacturer, catalog number, and magnification | In principle, easy to automatically capture, log and maintain during acquisition; in practice, varies significantly based on microscope manufacturer and image file format | Example: OME; 4DN-BINA-OME (NBO) Microscopy Metadata specifications # |
| **Image Acquisition Settings** | Describes the specific settings employed during image acquisition | Illumination power settings, exposure time, detector gain | | |
| **Image Structure** | Structure of the image data file | Pixel size, number of focal planes, channels, and time points, dimension order | | |

**Legend:** Microscopy Metadata documents the process of Image Acquisition using a Light Microscope. This table describes the different categories of Data Provenance Metadata that belong to Microscopy Metadata. # - (1–3, 10, 17, 85).

# Table III

## Table III: Microscopy Metadata - Quality Control Metadata

| Category | Metric | Description |
|---|---|---|
| Optical | Lateral resolution | Spatial resolving power achieved in x and y. According to ISO 21073:2019, Lateral resolution is defined by the FWHM of the intensity signal along a lateral direction through the centre of a fluorescent Point-like object (i.e., Spot or Point Emitter) positioned at the center of the FOV |
| Optical | Axial resolution | Spatial resolving power achieved in z. According to ISO 21073:2019, Axial resolution is defined by the FWHM of the intensity signal along the axial direction through the centre of a fluorescent Point-like object (i.e., Spot or Point Emitter) positioned at the center of the FOV |
| Optical | Planarity | Deformations of the focal plane across the FOV, due to misalignment/malfunction/objective aberrations |
| Optical | Lateral asymmetry | Deviation from lateral symmetry (i.e., along the x or y dimension) of the PSF as observed across the FOV |
| Optical | Axial assymmetry | Deviation from axial symmetry (i.e., along the z dimension; also referred to as spherical aberration) of the PSF as observed across the FOV |
| Optical | Lateral chromatic shift | Offset between color channels (and variance) in x and y |
| Optical | Axial chromatic shift | Offset between color channels (and variance) in z |
| Optical | Field flatness | Response variations over the FOV, due to unevenness across the illumination field or detector sensitivity and/or distortions in the optical path |
| Intensity/Excitation | Excitation wavelength | The center wavelength of the excitation light |
| Intensity/Excitation | Illumination power at the sample | Illumination power measured at the sample plane |
| Intensity/Excitation | Illumination power at back of objective | Illumination power measured at the back focal plane of the objective |
| Intensity/Excitation | Irradiance at the sample | Observed mean excitation light irradiance (radiant power per unit area) as measured at the sample plane over the FOV |
| Intensity/Excitation | Irradiance variance | Excitation light irradiance variation over the FOV |
| Intensity/Excitation | Illumination stability | Percentage deviation of illumination powers from the average during the time interval considered |
| Intensity/Excitation | Illumination linearity | Linearity of illumination response to different input illumination power settings, as measured at the sample |
| Intensity/Detector | Detection limit | Minimum detectable signal |
| Intensity/Detector | Dark offset signal | The average detected signal in the absence of incoming light |
| Intensity/Detector | Observed analog gain | Measured conversion factor between arbitrary digital units (a.u.) of intensity and photons, reported in terms of electrons/ADU (analog-to-digital unit) |
| Intensity/Detector | Linear range | Range of illumination intensity in which the response of the detector is linear |
| Intensity/Detector | Observed dark signal non-uniformity | Dark signal non-uniformity (DSNU) refers to the non-uniformity of response characteristics between pixels. DSNU is the standard deviation from the average pixel intensity detected across the imaging array at a particular setting (temperature, integration time) and in the absence of incoming light (useful to measure for sCMOS cameras) |
| Intensity/Detector | Observed dark noise | Dark noise arises from fluctuations in the thermal generation of electrons by the detector chip in the absence of light. Dark noise is measured in the absence of incoming light and grows with exposure time (useful to measure for long acquistion times) |
| Intensity/Detector | Observed read-out noise | Read-out noise is the inherent noise of the detector and arises during the read-out process when electrons are subjected to analog to digital conversion. Read-out noise is measured in the absence of incoming light and with minimal integration time. It does not depend upon exposure time and it is generally expressed in photo-electrons (useful to measure for low signal applications) |
| Mechanical | Lateral reproducibility | Precision with which the system is capable of maintaining a given (x, y) position over time. This metric might also be called lateral drift (useful for the calibration of sample positioning devices) |
| Mechanical | Focal reproducibility | Precision with which the system is capable of maintaining a given z position over time. This metric might also be called focal drift (useful for the calibration of sample positioning devices) |
| Mechanical | Lateral repeatability | Accuracy with which the system is capable of returning to a previously visited (x, y) position (useful for multipoint acquisition) |
| Mechanical | Focal repeatability | Accuracy with which the system is capable of returning to a previously visited z position (useful for multipoint acquisition) |
| Mechanical | Lateral settling time | The time it takes for the system to stabilize at a given target (x, y) position (useful for fast three-dimensional acquisitions) |
| Mechanical | Focal settling time | The time it takes for the system to stabilize at a given target z position (useful for fast three-dimensional acquisitions) |

**Legend:** Microscopy Metadata documents the process of Image Acquisition using a Light Microscope. This table describes suggested Quality Control metadata keys that belong to Microscopy Metadata. Abbreviations: FOV, Field-of-View; PSF, Point Spread Function